\documentclass[3p,times]{elsarticle}

\usepackage{ecrc}

\volume{00}

\firstpage{1}

\journalname{Nuclear Physics A}

\runauth{R. Seto, for the PHENIX Collaboration}

\jid{nupha}

\usepackage{amssymb}

\usepackage[figuresright]{rotating}
\usepackage{xspace}
\newcommand{\gone}{\mbox{G$^{(1)}$}\xspace}
\newcommand{\gtwo}{\mbox{G$^{(2)}$}\xspace}
\newcommand{\ww}{\mbox{Weizs\"{a}cker-Williams}\xspace}
\newcommand{\qsat}{\mbox{Q$_{sat}$}\xspace}

\newcommand {\auau}{\mbox{Au$+$Au}\xspace}

\newcommand {\pau}{\mbox{p$+$Au}\xspace}
\newcommand {\pa}{\mbox{p$+$A}\xspace}
\newcommand{\pt}{\mbox{${p_T}$}\xspace}

\newcommand {\ppau}{\mbox{p$^{\uparrow}+$Au}\xspace}
\newcommand {\ppp}{\mbox{p$^{\uparrow}+$p}\xspace}
\newcommand {\gdir}{\mbox{$\gamma_{\rm{direct}}$}\xspace}

\begin{document}

\begin{frontmatter}


\author{Richard Seto for the PHENIX Collaboration}
\ead{richard.seto@ucr.edu}
\address{Department of Physics and Astronomy, University of California, Riverside 92521}

\dochead{}

\title{The sPHENIX Forward Angle Spectrometer}


\author{}

\address{}

\begin{abstract}
sPHENIX is a major upgrade proposed for the PHENIX detector.  As part
of this proposal, a forward spectrometer to cover the rapidity range
$1<\eta<4$ is in a conceptual stage to complement the detectors in the
central region. The forward rapidity region is important for the study
of high density of gluons in a nucleus, as well as for the measurement
of the spin structure of the nucleon. Some of the basic physics goals
and a possible detector configuration are outlined.

\end{abstract}

\begin{keyword}
QCD, CGC, sQGP, RHIC, PHENIX

\end{keyword}

\end{frontmatter}



\section{Introduction}
The sPHENIX detector, proposed in~\cite{Aidala:2012sPHENIX} is the
first stage of a major upgrade to the PHENIX detector, which replaces
the present central magnet with a solenoid, thereby removing the large
iron yoke at forward rapidity that acts as a hadron absorber for
the current muon detectors. This will allow for the addition of a
forward spectrometer, covering $1<\eta<4$, with the capability of
measuring hadrons, electrons, photons and jets. One of the primary goals
of the forward sPHENIX spectrometer (fsPHENIX) is the study of cold
nuclear matter (CNM) in proton-nucleus collisions. Additional goals
include the study of spin asymmetries in transversely polarized \ppp
collisions as well as the possibility of making measurements in heavy ion
collisions. The design is such that sPHENIX could naturally be evolved
into a detector for the study of electron-proton and electron-nucleus
collisions, with the advent of a future high intensity electron beam
at RHIC.

\section{Gluon Saturation and Cold Nuclear Matter}

One of the main goals of the sPHENIX forward spectrometer is to understand the
dynamics of partons, primarily gluons, at very small momentum fraction
($x$) and high density. The gluon distributions in a nucleon rapidly
increase as one goes to low-x. Because of the uncertainly principle,
however, they also increase in transverse size and begin to overlap
and recombine, resulting in saturation. In the case of a nucleus, 
this effect is magnified by a factor A$^{1/3}$ reflecting the nuclear thickness. 
One of the most successful descriptions of this phenomenon is the Color-Glass
Condensate (CGC)~\cite{Gelis:2010nm}.  Two hints of this saturated
state of gluons are the suppression of forward hadron production
and the disappearance of the correlations of back-to-back hadrons in
\pa collisions. As in the case of many other phenomena related to
QCD (e.g. jet triplicity as a signature of gluons and the correctness
of QCD as a theory, or flow and suppression of high \pt hadrons as a
signature of the sQGP), it appears as if there may be no incontrovertible
signature of the CGC~\cite{Albacete:HP2012}.  There are other seemingly
competing descriptions, namely transverse momentum dependent (TMD)
parton distribution functions (PDFs), and higher twist
formulations~\cite{Qiu:2003vd} of shadowing. Recently, however, the
CGC and the TMD approaches have been shown to be equivalent for
certain kinematical ranges~\cite{Dominguez:2011wm}.

The goals of the forward sPHENIX program will be to measure the
parameters relevant to the CGC, namely the saturation scale \qsat, and
the predicted gluon distributions which manifest themselves in
distributions of hadrons, photons and jets. As with other advances
made in QCD, a preponderance of measurements consistent with the
theoretical expectations will serve to substantiate the basic
ideas. There are two gluons distributions of importance, the \ww
distribution, \gone, and the dipole distribution,
\gtwo~\cite{Dominguez:2011wm}. Single particle observables can only
give a limited amount of information. However, for direct
photon(\gdir)+jet and dijet final states, one can calculate an
effective parton momentum fraction x, given by the usual relationship
between x and two measurable quantities, rapidity and \pt. The
\gdir+jet and dijet final states in \pa collisions have different
sensitivities to the gluon distributions \gone and \gtwo. The
\gdir+jet process depends only on \gtwo, while dijet processes are
dependent on both \gone and \gtwo. One can begin by making
measurements of \gdir+jet to determine \gtwo, then by making
measurements of dijet events and knowing \gtwo, one can extract
\gone. In addition, \gdir+jet events can also be used to make a
measurement of \qsat. Recent work by Jalilian-Marian and
Resaein~\cite{JalilianMarian:2012bd} shows that the strength of the
correlation between a direct photon and an opposing hadron is
sensitive to the saturation scale.  The correlation between virtual
photons, as measured by Drell-Yan pairs, and hadrons has also been
studied~\cite{Stasto:2012ru}.  While the latter is a very difficult
measurement because of the low cross section which would yield only
several thousand pairs in a typical 20 week run of \pau, we are
exploring the possibility of colliding protons on lighter ions since
increase in luminosity more than compensates for the fact that there
will be fewer binary collisions, although the effects due to
saturation may also be correspondingly smaller. All of these detailed
measurements will take time and require a detector with good photon
identification, reasonable jet resolution and good acceptance. Once
the data is taken and analyzed, however, one can assemble the results
into a complete picture, which should indicate whether the basic
tenets of the CGC model or other models are correct.

An intriguing new signature utilizing polarized proton+nucleus (\ppau)
collisions has been suggested~\cite{Kang:2011ni}.  The single
transverse spin asymmetry A$_N$, is defined for a forward moving
vertically polarized proton beam, which scatters with its products to
the left of the beam, as
$$A_N \equiv
\frac{\sigma^\uparrow-\sigma^\downarrow}{\sigma^\uparrow+\sigma^\downarrow},$$
where the arrows indicate the direction of beam polarization.  The
authors indicate that the strength of A$_N$ measured in \ppau
collisions relative to transversely polarized p+p (\ppp) collisions
would be directly related to the saturation scale. The most
straightforward reaction would have a charged or neutral pion in the
final state, where the asymmetry is known to be large in \ppp
collisions at forward rapidity~\cite{Adams:2003fx}. Such a measurement
is unique to the RHIC complex, the only machine in the world which can
deliver polarized proton+nucleus collisions.  These measurements
will, of course, be extended to other final state particles, e.g. \gdir,
jets, heavy quarks, and quarkonia.

\section{Other Physics}

The goals of the sPHENIX forward spectrometer, encompass more than the study
of gluon saturation effects. These other goals have significant
influence on the detector design.  Additional studies in CNM will
include studying (a) the mechanisms that cause transverse momentum
broadening and energy loss of partons in CNM and (b) hadronization
mechanisms and time scales and their modification in a nuclear
environment. In addition to the study of nuclear matter, there will be
a significant program to study the origin of nucleon spin. Two of the
signatures of major importance are Drell-Yan pairs and jets.  The high
luminosity available for 500 GeV center of mass collisions, the large
acceptance of the forward arms, and long running periods of 20
weeks, will yield enough DY pairs for a significant measurement.  We
are also exploring the role fsPHENIX could have in studying Au+Au
collisions.  One of the key questions that might be probed is the
mechanism of equilibration and formation of the sQGP, since the
forward arms together with the central barrel will have very large
rapidity coverage for the study of long range correlations. fsPHENIX
will also have the capability to study the sQGP under a variety of
densities, since the rapidity plateau drops to about half of its
maximum in the acceptance of the forward arms at RHIC energies.

\section{Detector Considerations}

\begin{figure}
\centering
\includegraphics[width=0.8\linewidth]{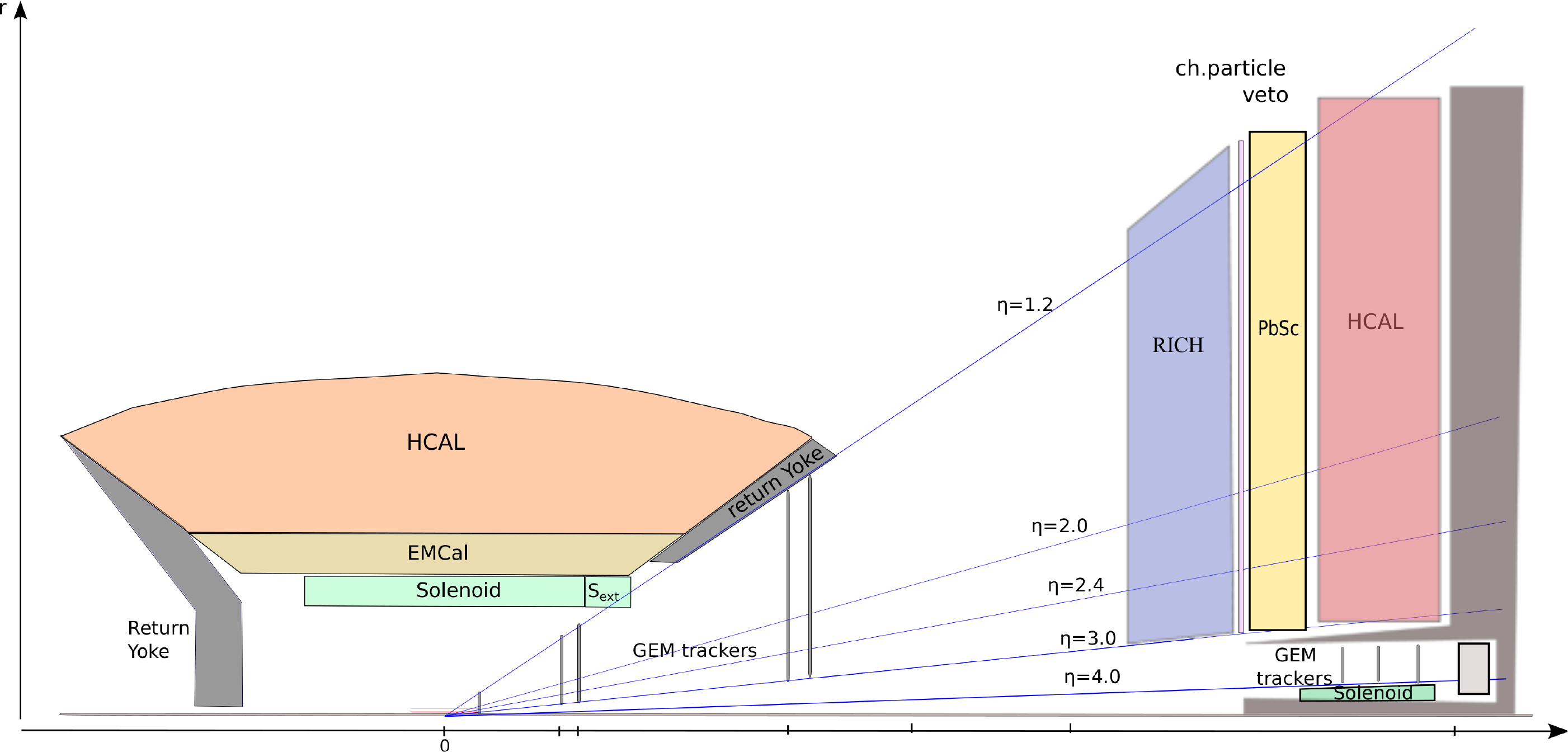}
\caption{\label{fig:detector} A straw man layout of a possible detector for a future forward upgrade to sPHENIX.}
\end{figure}

The broad range of physics topics to be explored by fsPHENIX, require
a detector similar to the central barrel, capable of measuring photons
and jets. In addition, Drell-Yan measurements for the spin program
require excellent dilepton identification and the ability to reduce
backgrounds from correlated charm and beauty decays. Both the CNM and
spin programs require vertexing capability to identify heavy
quarks. There are two significant differences between the requirements
for the forward arms as compared to the central barrel.  The first is
that very forward rapidity measurements must be done at small
angles to the beam, where the density of particles is high, and
momenta are large.  The second is the need for hadron particle
identification, primarily for spin physics, to avoid cancellation of
transverse spin effects in the fragmentation of different hadrons,
since these effects are dependent on the flavors of the quarks
involved in the process.

The forward sPHENIX study group has been investigating detector
performance design requirements needed to meet the desired physics
goals. The envisioned fsPHENIX forward spectrometer will have an
acceptance from a pseudorapidity of 1.2 to 4.  The acceptances of the
mid-rapidity upgrade and the forward upgrade will be matched as
closely as possible in order maintain a uniform acceptance for jet
measurements.  Currently, a ``straw man'' design
(Fig.~\ref{fig:detector}) is being used for the purpose of sensitivity
studies. This design divides fsPHENIX into two sections.

The first section covers a region of $\eta$ from 1.2 to 3.0. In this
region, an extension or modification of the central solenoid provides a
sufficiently strong magnetic field for good momentum resolution. Gas
Electron Multiplier (GEM) detectors provide charged particle
tracking. Silicon detectors are located near the collision point, to
provide vertexing capability for heavy quarks.  Particle
identification is based on a Ring Imaging Cerenkov Detector, and will
probably be usable only in the low multiplicity environment of \ppp
collisions, though it may also be useful in \pa collisions. The forward
electromagnetic calorimeter consists of a reconfiguration of the current
PHENIX electromagnetic calorimeters (EMCal) and would be followed by a
hadronic calorimeter with modest energy resolution (HCAL). The front
face of the EMCal would be located between 3 and 5 meters from the
interaction point.  The HCAL together with the EMCal provide
sufficient jet resolution for use in \ppp and \pa collisions. The
capability of the spectrometer to measure jets in
\auau collisions is under study, both in terms of physics reach and
detector capability. A charged particle veto or pre-shower detector
would be located in front of the EMCal to aid in photon
identification.  Studies are underway to determine the feasibility of
tracking muons with a muon identifier behind the HCAL, which would
allow for di-lepton measurements to be made in both the electron and
muon channels.

In the very forward section, from $\eta$ = 3 to 4, an additional source
of magnetic field would be needed to retain the momentum
resolution. Currently, the design calls for an radial magnetic field with
tracking provided by GEM detectors located in the magnet. A reconfiguration of
the current lead tungstate Muon Piston Calorimeter, which provides the 
measurement of electromagnetic showers, is located behind the tracking
detectors. The possibility of adding hadronic calorimetry in this region
is currently being discussed. It must be emphasized
that this design is undergoing a process of significant
evolution, but the present ``straw man'' design provides the basis
of studies to better define and demonstrate the capability of the
fsPHENIX detector to address the physics goals outlined.

\section{Conclusion}

\begin{figure}[hbt]
\centering
\includegraphics[width=0.4\linewidth]{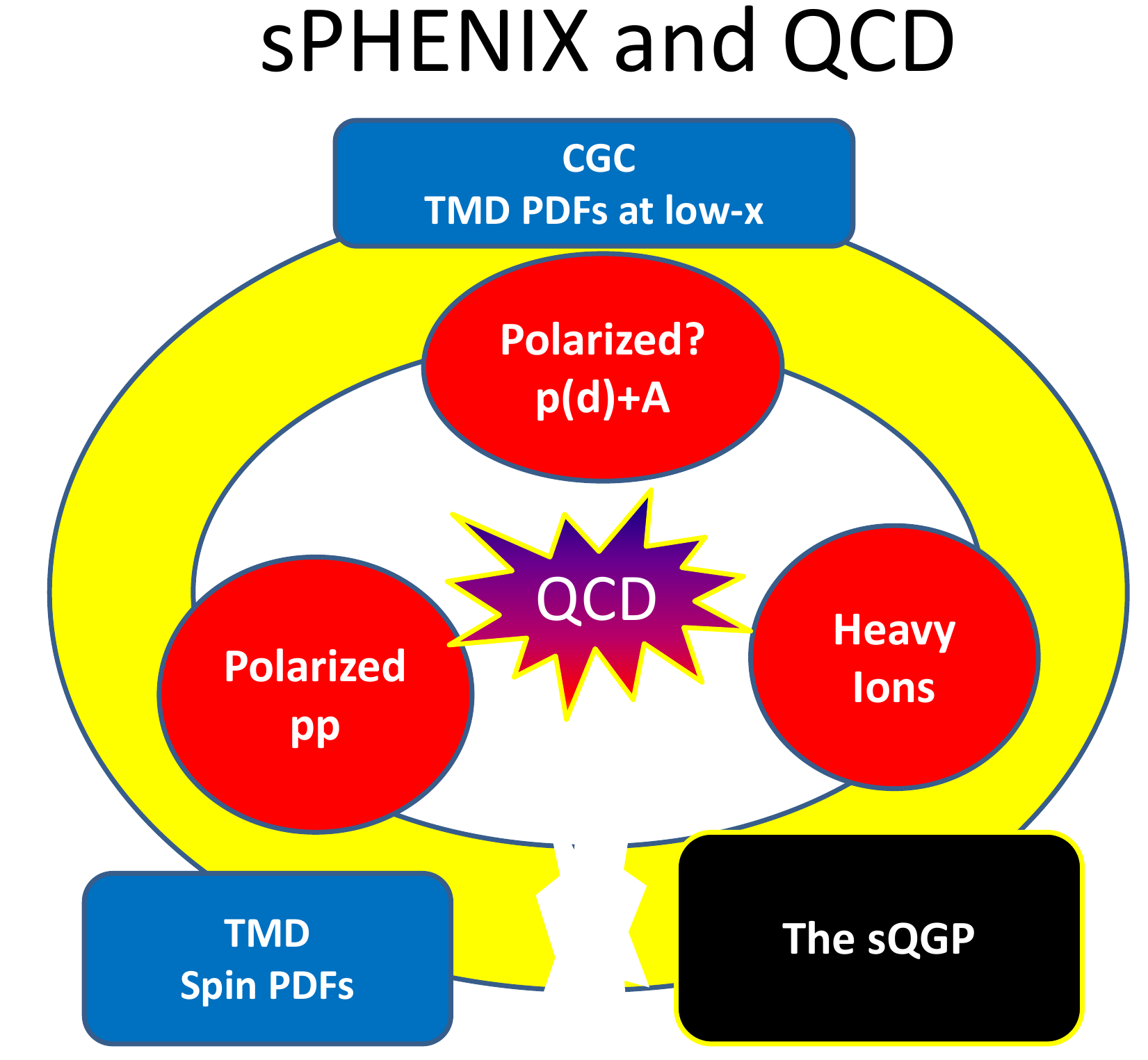}
\caption{\label{fig:QCD}A schematic diagram to emphasize the fact that
three areas of QCD study which have seemed disparate in the past, actually
have a close relationship, both in the theoretical techniques used to
understand them and in the interconnectedness of the phenomena exemplified by them.}
\end{figure}

While the precise physics signatures and detector design are still in
a conceptual stage, the basic physics goals are clear. They are: (a)
to elucidate the nature of gluons at small
x in a nucleus, and to measure the relevant parameters and
distributions such as \qsat and the \ww and dipole gluon
distributions, both for an intrinsic understanding of the saturation of gluon
density, and
as the initial state of the sQGP; (b) to understand the nature of nucleon
transverse spin, and its connection to orbital angular momentum; (c)
to understand the early stages of the formation of the sQGP. One of
the highlights of recent developments in theory and experiment, is the
extent to which the three areas of relativistic heavy ion collisions,
the study \pa collisions and Cold Nuclear Matter, and the spin physics
of the nucleon, are interrelated (Fig.~\ref{fig:QCD}).  Transverse
momentum dependent methods first used in the study of nucleon spin
have also given insight into the saturation of gluon density. In fact 
transversely
polarized \ppau collisions may be a means of measuring \qsat. In turn,
high density gluons as described by a Color Glass Condensate model are
probably the best candidate as the initial state for the strongly
interacting Quark Gluons Plasma.  The understanding of the strong
interaction is an exciting field, with phenomena as rich and varied as
manifestations of the electromagnetic interaction as exemplified by
the fields of atomic and condensed matter physics.




\bibliographystyle{elsarticle-num}
\bibliography{<your-bib-database>}



\end{document}